\documentclass{article}

\def\a{\alpha}

\def\be{\beta}
\def\g{\gamma}
\def\ga{\gamma}
\def\de{\delta}   
\def\eps{\varepsilon}
\def\la{\lambda}

\def\th{\theta}

\def\vphi{\varphi}
\def\De{\Delta}
\def\La{\Lambda}
\def\phi{\varphi}

\def\E{{\mathcal E}}
\def\R{{\bf R}}
\def\L{{\mathcal L}}
\def\T{{\rm T}}
\def\G{{\mathcal G}}
\def\J{{\mathcal P}}
\def\P{{\mathcal P}}
\def\B{{\mathcal B}}
\def\M{{\mathcal M}}
\def\Jb{{\bf P}}

\def\Div{{\tt Div }}
\def\Tr{{\tt Tr}}

\def\pa{\partial}
\def\pd{\partial}
\def\d{{\rm d}}       
\def\grad{\nabla}     
\def\na{\nabla}
\def\ss{\subset}
\def\sse{\subseteq}
\def\unm{{1 \over 2}}

\def\EOP{\hfill $\diamondsuit$}

\def\({\left(}
\def\){\right)}
\def\[{\left[}
\def\]{\right]}
\def\~#1{\widetilde #1}
\def\.#1{\dot #1}
\def\^#1{\widehat #1}

\def\quotes#1{``#1''}

\def\beq{\begin{equation}}
\def\eeq{\end{equation}}
\def\lab#1{\label{#1}}

\def\qd{{\dot q}}

\def \ov{\over}

\def \lb{\label}

\def\={\, =\, }

\def\dst{\displaystyle}

\def \sy {symmetry}
\def \sys {symmetries}
\def \so {solution}
\def \eq{equation}
\def\LG{Lagrangian}

\def \q{\quad}

\def \sk{\medskip}
\def \ni{\noindent}

\def \ls{$\la$-symmetry}
\def \ms{$\mu$-symmetry}

\def \li {{\pd \L\ov{\pd u_i^a}}}
\def \lij {{\pd \L\ov{\pd u^a_{ij}}}}
\def\vf{vector field}

\begin{document}

\title{Noether theorem for $\mu$-symmetries}

\author{Giampaolo Cicogna\\ {\it Dipartimento di Fisica, Universit\`a di
Pisa}  \\
and {\it INFN, Sezione di Pisa,} \\
{\it Largo B. Pontecorvo 3, 50127 Pisa (Italy)} \\ {\tt
cicogna@df.unipi.it}
\\ {} \\ Giuseppe Gaeta\\ {\it Dipartimento di Matematica,
Universit\`a di Milano,} \\ {\it via Saldini 50, 20133 Milano
(Italy)} \\ {\tt gaeta@mat.unimi.it}}

\date{}

\maketitle

\noindent {\bf Summary.} We give a version of Noether theorem
adapted to the framework of $\mu$-symmetries; this
extends to such case recent work by Muriel, Romero and Olver
in the framework of $\lambda$-symmetries, and connects
$\mu$-symmetries of a Lagrangian to a suitably modified
conservation law. In some cases this ``$\mu$-conservation law''
actually reduces to a standard one; we also note a relation
between $\mu$-symmetries and  conditional invariants. We also
consider the case where the variational principle is itself
formulated as requiring vanishing variation under $\mu$-prolonged
variation fields, leading to modified Euler-Lagrange equations. In
this setting $\mu$-symmetries of the Lagrangian correspond to
standard conservation laws as in the standard Noether theorem. We
finally propose some applications and examples.

\section*{Introduction}

The study of
nonlinear differential equations was the main motivation to Sophus
Lie when he created what is nowadays known as the theory of Lie
groups and Lie algebras. After a period of near-oblivion, symmetry
methods are now recognized as one of the most effective tools for
the study of differential equations -- both in geometrical sense
and for what concerns the search for explicit solutions
\cite{Gbook,KrV,Olv1,Ste}.

The relevance of symmetry properties in Analytical Mechanics and
Field Theory underwent a similar parabola; after Emmy Noether
established her theorem \cite{Noe}, it has somehow been overlooked
for some time, and then came to be recognized as a tremendously
important tool, actually at the basis of all the conservation laws
of Physics (see \cite{Kos} for an history of the appraisal of
Noether theorem, as well as for an in-depth discussion of it). It
is thus not surprising at all that Noether theorem is a central
issue not only in treatments of Analytical Mechanics and Field
Theory \cite{Arn,LLmec,LLft}, but also  in the treatment of
symmetry methods for differential equations, in particular when
one is concerned with equations arising from a variational
principle \cite{Olv1}.

The effectiveness of symmetry methods for differential equations
led various authors to consider generalizations in several
directions (which we will not discuss in general). Here we are
specially interested in one of these, first proposed by M.C.
Muriel and J.L. Romero under the name of ``$C^\infty
(M^{(1)})$-symmetries'' or ``$C^\infty$-symmetries'' for short, or
finally of ``$\la$-symmetries'', in the framework of ODEs
\cite{MuRom1}, and then generalized to PDEs under the name of
``$\mu$-symmetries'' \cite{CGM,Gspt,GM}.

It should be stressed that in this case the generalization with
respect to standard Lie symmetries lies not in the definition of
``symmetry'', but in the way vector fields are prolonged from the
phase manifold $\M$ (the manifold of dependent and independent
variables, see below) to jet manifolds $J^r \M$ of appropriate
order for the equation at hand. In this, they are different from
other proposed generalizations of Lie symmetries.

Definition and properties of $\mu$-prolongations and symmetries
(which include $\la$-ones as a special case) are recalled and
discussed in section \ref{sect:mu} below, so here we just recall
that these allow to integrate by symmetry methods equations which
are known not to admit any standard Lie symmetry
\cite{MuRom1,MuRom2}; needless to say, this is a major reason (but
not the only one) for the interest they raised.

It is then natural to wonder if Noether theorem can be somehow
extended\footnote{The original Noether theorem was actually
already much more general than the simple version usually met in
Mechanics  textbooks \cite{Arn,LLmec}, as discussed at length in
\cite{Kos,Olv1}. Several extensions of Noether theorem have also
been given in the literature, see e.g. \cite{CaS,Kos,Olv1}. Here
we limit our discussion to the simpler forms of the Noether
theorem.} to encompass $\la$- or $\mu$-symmetries as well.
This question was tackled by Muriel, Romero and Olver \cite{MRO},
who were able to give an adapted formulation of Noether theorem
relating $\la$-symmetries of the Lagrangian $\L$ defining the
variational problem and conservation laws in a suitably modified
sense for problems giving raise to ODEs -- the latter being of
course the Euler-Lagrange equations for $\L$.

In the present note we have a twofold aim. On the one hand, we
extend the results obtained by Muriel, Romero and Olver in the
case of $\la$-symmetries -- hence Mechanics and ODEs -- to the
case of $\mu$-symmetries and hence of Field Theory and PDEs. In
this way we show that a $\mu$-symmetry of the Lagrangian leads to
a modified ``conservation law'' -- called a {\it
$\mu$-conservation law} -- under the evolution law described by
the standard Euler-Lagrange equations for $\L$. In some cases, the
$\mu$-conservation law actually reduces to a standard one. 
We also note a relation between $\mu$-symmetries
and {\it conditional invariants}.

On the other hand, we want also to raise a different question.
That is, the Euler-Lagrange equations are obtained under the
assumption that given a variation $\de u$ of the dependent
variables $u$, partial derivatives of these are varied
accordingly; this corresponds to considering a ``variation field''
$\Delta$ (in the simplest case, $\Delta = \phi^a (u) (\pa / \pa
u^a)$) acting in the phase manifold $\M$, and lifting it to the
first jet space $J^1 \M$ -- or to higher order jet space for
Lagrangians of order higher than one -- according to the standard
prolongation law. The requirement of zero variation for the action
integral leads then, in the familiar way, to establishing the
Euler-Lagrange equations.

But, as recalled above, in $\la$- and $\mu$-symmetries one is
actually modifying the prolongation operation; it is then rather
natural to consider what happens when the requirement of zero
variation for the action integral is considered under variation
vector fields which are lifted to jet manifolds with the modified
prolongation operation. In this case one obtains a modified
version of the Euler-Lagrange equations -- which will be
christened as the $\mu$-Euler-Lagrange equations -- and hence a
{\it modified} evolution law. In the final part of the present
paper, we show that when we consider this modified evolution law,
a $\mu$-symmetry of the Lagrangian leads to a standard
conservation law. This will also be interpreted in the light of
the findings of \cite{CGM}.

We choose to mostly limit our discussion to first order
Lagrangians; this case -- beside being of special physical
interest -- allows a simpler notation and does not hide behind
complicate formulas the main point of our discussion, which would
extend to higher order Lagrangians in the same way as done for the
standard Noether theorem.  Similarly, our examples are as simple
as possible in order to illustrate the results and the main points
of our discussion.

The paper is organized as follows: in sect.1 we establish some
general notation; in sect.2 we recall the basics of
$\mu$-prolongation and $\mu$-symmetries, and define
$\mu$-conservation laws. In sect.3 we establish a version of
Noether theorem for $\mu$-symmetries; this yields correspondence
with a $\mu$-conservation law. In sect.4 we extend this result to
the case of divergence $\mu$-symmetries. In sect.5 we establish
the connection between $\mu$-symmetries and (standard) conditional
invariants. Sect.6 is devoted to recalling the gauge equivalence
between $\mu$-prolongation and standard ones; this allow to
establish that in a number of cases -- but not in general ! -- the
$\mu$-conservation law associated to a $\mu$-symmetry of the
Lagrangian is actually a standard conservation law. In sect.7 we
establish the $\mu$-Euler-Lagrange equations for a Lagrangian
$\L$, and show that a $\mu$-symmetry of $\L$ yields a standard
conservation law under such an evolution law; this result is also
seen in relation with the gauge equivalence mentioned above.
Section  8 is devoted to several illustrative examples
and  applications; some short conclusions will be presented in sect.9.

\bigskip

\noindent {\bf Acknowledgements.} The research of GG was partially
supported by MUR (Italy) under program PRIN-COFIN2004, project
MMxDD. We thank G. Saccomandi for reviving our interest in the
relations between Noether theorem and $\mu$-symmetries, and P.
Morando for useful discussions, in particular concerning the
results of section \ref{sect:muEL}. 


\section{General notation}

We will preliminarly set up some general notation, to be used as a
default in the following \cite{Gbook,KrV,Olv1,Ste}.

We will denote independent variables as $x = (x^1,...,x^p) \in B
\sse \R^p$; similarly we denote dependent variables (fields) as $u
= (u^1,...,u^q) \in U \sse \R^q$. The total manifold of dependent
and independent variables will be denoted as $\M = B \times U$; it
has the structure of a fiber bundle over $B$, and one will also
(implicitly) consider the associated jet bundles $J^k \M$. In the
following we will customarily assume that actually $B = \R^p$, $U
= \R^q$ (and hence $\M$ is also a linear space), as this will
simplify some notation; the discussion could  dispense with
this assumption with no extra difficulty but a heavier notation in
sections \ref{subsect:intermediate} and~\ref{subsect:gaugeequiv}. 

We use indices $i,j,k...$ for independent variables, and $a,b,c...$ for 
dependent ones (thus denoted respectively e.g. by $x^i$ and $u^a$). 
Sum over 
repeated indices (and multi-indices, see below) is always understood. The
partial derivative of $u^a$ with respect to $x^i$ will be denoted
in shorthand notation as $u^a_i$.
Total derivative operators will be denoted as $D_i$.
We employ whenever appropriate multi-index notation; a generic
multi-index will be denoted as $J = (j_1,...,j_p)$, and its order
by $|J|:= j_1 + ... + j_p$. By $u^a_J$ one means $(\pa^{|J|} u^a /
(\pa^{j_1} x^1 ... \pa^{j_p} x^p) )$; by $D_J$ one means
$D_i^{j_1} ... D_p^{j_p}$. 

Vector fields in $\M$ will usually be written (in coordinates) as
\beq\lb{1} X \ = \ \xi^i (x,u) \, {\pa \over \pa x^i} \ + \ \phi^a
(x,u) \, {\pa \over \pa u^a} \ . \eeq 
The $r$-th order (standard) prolongation \cite{Gbook,KrV,Olv1,Ste} 
of such a vector field will be usually written as
(here the sum is over all multi-indices with $|J| \le r$)
\beq\lb{2} X^{(r)} \ = \ \xi^i (x,u) \, {\pa \over \pa x^i} \ + \
\phi^a_J (x,u) \, {\pa \over \pa u^a_J} \ ; \eeq 
the coefficients
$\phi^a_J$ satisfy $\phi^a_0 = \phi^a$ and the  standard
prolongation formula 
\beq\lb{3}
\phi^a_{J,k} \ = \ D_k \phi^a_J \, - \, u^a_{J,i} D_k \xi^i \ .
\eeq
We will deal with Lagrangians; albeit these could depend on $x^i$, we 
will usually
ignore this dependence, and write $\L = \L (u,u_x)$ when we want to stress 
these are first order Lagrangians. The momenta are generally denoted by 
$\pi$, with
$$ \pi^i_a \ := \ {\pa \L \over \pa u^a_i} \ . $$

Finally, several of the objects we consider will take value in a set of 
matrices
(usually corresponding to the representation of a Lie algebra or group); we
denote this by saying they are M-functions. Thus a M-vector will be a 
vector
whose component take value on a space of matrices.

\section{$\mu$-prolongations, $\mu$-symmetries,
\\ and $\mu$-con\-ser\-va\-tion laws}
\label{sect:mu}

We will first give general definitions for $\mu$-prolongations and
$\mu$-symmetries. Later on, in section  6, we briefly recall their
relation with gauge transformations.

We refer to \cite{CGM,Gspt,GM} for further details on
$\mu$-prolongations and $\mu$-sym\-metries. These include as special
cases the $\la$-prolongations and $\la$-symmetries; see
\cite{MuRom1,MuRom2,MuVigo,MRO,PuS} for further detail on these.

\subsection{$\mu$-prolongations}

In $\mu$-prolongations, one equips $\M$ with a semi-basic form
\beq\lab{eq:mu} \mu \ = \ \La_i (x,u,u_x) \, \d x^i \eeq with
$\La_i$ a function taking values in a representation $T_\G$ of a
Lie algebra $\G$ acting in $F$. (In the following we will identify
the representation and the Lie algebra for ease of language, hence
call $\mu$ a $\G$-valued form rather than a $T_\G$-valued one.)

It should be stressed that the $\La_i$ ($i=1,...,p$) are square
$(q\times q)$ matrices; with the notation introduced above, we
also say that they are M-functions of the basic dependent
variables $u$ and  independent ones $x$, and of the first
derivatives~$u_x$.

For $p=1$ and $\La = \la I$ (with $\la (x,u,u_x)$ a scalar
function and $I$ the identity matrix) the M-function $\La$
actually corresponds to a standard scalar function $\la$, and we
recover the setting of $\la$-symmetries \cite{MuRom1}. Note that we get
a very similar setting even in the case where $p > 1$, but the
$\La_i$ are of the form $\La_i = \la_i I$, with $\la_i = \la_i
(x,u,u_x)$ scalar functions \cite{MuVigo}.

The $\G$-valued functions $\La_i$ should satisfy the compatibility
condition \beq\lab{eq:comp_cond} D_i \, \La_j \ - \ D_j \, \La_i \
+ \ [\La_i , \La_j ] \ = \ 0 \ ; \eeq where $D_i$ is the total
derivative with respect to $x_i$; equivalently (but in
coordi\-na\-tes-free notation), the $\G$-valued form $\mu$ must
satisfy the {\it horizontal Maurer-Cartan equation}
\beq\lab{eq:hMC} D \, \mu \ + \unm \ [\mu , \mu ] \ = \ 0 \ .
\eeq
The $\mu$-prolonged vector field acting in $J^{(r)} \M$ will be
denoted as $X^{(r)}_{(\mu)}$, i.e. \beq X^{(r)}_{(\mu)} \= \xi^i
{\pa \over \pa x^i} \ + \ \psi^a_J{\pa\over{\pa u^a_J}} \eeq 
where sum over repeated indices and derivation multi-indices (of order
$|J|\le r$) is implied, and where
the familiar prolongation formula (\ref{3}) is now replaced by
\beq\lab{eq:muprol0_rec}
\psi^a_{J,k} \ = \ D_k \psi^a_J \ - \ u^a_{J,i} \, D_k \xi^i \ + \
(\La_k)^a_{\ b} \( \psi^b_J - u^b_{J,i} \xi^i \)   \eeq
with $\psi^a_0=\phi^a$. This is also rewritten as
\beq\lab{eq:muprol_rec} \psi^a_{J,k} \ = \ \grad_k \psi^a_J \ - \
u^a_{J,i} \, \grad_k \xi^i \ , \eeq where we have introduced the
$\G$-valued differential operators 
\beq\label{eq:nabla} \grad_k \
:= \ \de^a_{\ b} D_k + (\La_k)^a_{\ b} \ . \eeq 
The operators
$\grad_i$ will play a key role in our constructions. We stress
that for $\La_i = 0$ the $\grad_i$ reduce to the familiar total
derivative operators $D_i$; in general, the $\grad_i$ take the
place of the $D_i$ when we consider $\mu$-prolongations (and
related concepts) instead of standard ones.

Note that the compatibility condition (\ref{eq:comp_cond}) is
written, with the notation (\ref{eq:nabla}), as the zero-curvature
condition \beq [ \grad_i , \grad_j ] \ = \ 0 \ . \eeq

Formulas become simpler when we deal with {\it
evolutionary representatives} of vector fields; these are vertical
for the fibration $(\M,\pi,B)$, and the evolutionary representative
of $X$ reads
\beq\lb{4} X_Q \ = \ Q^a
(x,u,u_x) {\pa \over \pa u^a} \ \ \ {\rm with} \q Q^a := \phi^a -
\xi^i \, u^a_i \ . \eeq
With this, the recursion formula (\ref{eq:muprol_rec}) for the 
$\mu$-prolongation
becomes, with obvious notation,
\beq\lab{eq:muprol_evo} Q^a_J \ = \ \grad_J Q^a \ . \eeq

\subsection{$\mu$-symmetries}

Given a system $\De$ of differential equations of order $r$, call
$S_\De$ the corresponding solution manifold in $J^r \M$. We say
that $X$ is a $\mu$-symmetry of the system $\De$ (for the given
form $\mu$) if $X^{(r)}_{(\mu)} : S_\De \to \T S_\De$.

We will often use $Y$ as a shorthand notation for $X^{(r)}_{(\mu)}$.

Similarly, given a function ${\mathcal F} : J^r \M \to \R$, we say
that $X$ is a $\mu$-symmetry for ${\mathcal F}$ (or equivalently
that ${\mathcal F}$ is $\mu$-invariant under $X$) if
$Y[{\mathcal F}] = 0$; this is also expressed as
the invariance of level manifolds of ${\mathcal F}$ under
$Y$.

We stress that -- as discussed in \cite{CGM} -- $\mu$-\sys\ are
not \sys\ in the usual sense; in particular, they do not in
general transform \so s of the differential problem into other \so
s. (Moreover, $\la$- and $\mu$-symmetries of a given equation do
not in general form a Lie algebra.) It was shown in
\cite{MuRom1,MuVigo} and then in \cite{CGM,GM} that they are
essentially as good as standard symmetries as far as determining
exact solutions of the differential problem is concerned. We will
show, generalizing some previous results \cite{MRO}, that they can
also be quite useful in the context of Noether theory.

In the following, we will deal with \vf s $X$ having $\xi^i= 0$, i.e.
$Q^a\equiv \phi^a$;
these represent transformations in the space of dependent
variables (fields) not affecting nor depending on independent
variables (note this is the same class of transformations
considered in the standard formulation of the classical Noether
theorem \cite{Arn,LLmec}). This assumption greatly simplifies the
formulas, in particular the expression of the invariance condition
of the Lagrangian, and of the conservation laws.

In Sect.~4 we will show how to extend our results to the case of divergence
\sys ;   as well known, see \cite{Olv1},   this can allow us
to include in our discussion not only \vf s with $\xi^i\not=0$
(introducing evolutionary \vf s), but also the case of generalized
\vf s.

\subsection{Reduction of $\mu$-prolongation to ordinary and $\la$
ones}
\label{subsect:intermediate}

We can write the coefficients $\psi^a_J$ of the
$\mu$-prolongation (\ref{eq:muprol0_rec}) as 
$$ \psi^a_J \ = \ \phi^a_J \, + \, F^a_J $$
where $\phi^a_J$ are the coefficients of the standard
prolongation (\ref{2},\ref{3}); inspection of the standard and
$\mu$-prolongation formulas in recursive form (see \cite{GM} for
details) shows that the recursion formula for the $\mu$-difference
terms $F^a_J$ is 
$$F^a_{J,i} \ = \ \[ \de^a_{\
b} \, D_i \, + \, (\La_i)^a_{\ b} \] \, F^b_J \ + \ (\La_i)^a_{\
b} \,  D_J Q^b  \ =\ \grad_i F^a_J + (\La_i)^a_{\ b} D_J Q^b $$
with of course $F^a_0 = 0$.
It follows, as discussed at length in
\cite{CGM}, that standard and $\mu$-prolon\-gations coincide on
vector fields such that the $Q^a$ belong identically (hence with
all their partial derivatives) to the null space of all the
matrices $\La_i$.

In particular, if the $\La_i$ belong to a representation $T_\G$ of
the Lie algebra $\G$, acting in the $M = \R^p \times \R^q$ space by a
linear and non-free action via this, and we call $M_0 \sse \T \M$ the
subspace fixed under all of $T_\G$, then all vector fields having only
components along $\T M_0$ will be in this class.

An interesting \quotes{intermediate} case arises when the $\La_i$ admit
common eigen\-spaces, in general with different eigenvalues for
different $\La_i$ (see also subsection \ref{subsect:gaugeinter} in this 
respect).
In fact, suppose there is a subspace $M_1$
such that $\La_i {\bf v} = \la_i {\bf v}$ for all $i = 1,...,p$
and all ${\bf v} \in \T M_1$. Then -- as long as we consider
vector fields such that $Q^a$ belongs identically to $\T M_1$, for
all $a=1,...,q$ -- things behave as if we were in the case $\La_i
= \la_i I$, i.e. by all means we deal with the case of \quotes{vector}
$\la$-prolongations and symmetries \cite{MuVigo}.

\subsection{$\mu$-conservation laws}

A (standard) {\it conservation law} is a relation
\beq\label{eq:standconslaw} D_i \Jb^i \ = \ 0 \ , \eeq 
where
$\Jb^i$ is a  $p$-dimensional vector.

In our case, we define a {\bf $\mu$-conservation law} as a relation 
\beq\label{eq:muconslaw0} \Tr \, \( \grad_i \, \J^i \) \
= \ 0 \eeq 
involving the $\G$-valued differential operators
$\grad_i$ and some $\G$-valued M-vector $\J$. 
In components, (\ref{eq:muconslaw0}) reads
\beq\label{eq:muconslaw} (\grad_i)^a_{\ b} \ (\J^i)^b_{\ a} \ \equiv \
(\grad^a_{\ b})_i \ (\J^b_{\ a})^i \ = \ 0 \ . \eeq Thus $\J^i$ is
a (matrix-valued) vector whose divergence with respect to the
(matrix) differential operator $\grad$ vanishes; note here the
divergence is defined with the help of the trace operation, see
(\ref{eq:muconslaw0}). In this case the M-vector $\J$ will be called
a {\bf $\mu$-conserved vector}.

Note that $\mu$-conservation does not imply conservation of $\J$
nor of its trace  in ordinary sense; if $\J$ is a $\mu$-conserved
vector, then -- putting now $\Jb^i=\Tr(\J^i)$ and recalling the
definition (\ref{eq:nabla}) -- it satisfies 
\beq\label{eq:nabJ0}
D_i\Jb^i\ \equiv \ D_i (\J^i)^a_{\ a} \= - \ (\La_i)^a_{\ b}
(\J^i)^b_{\ a} \=-\,\Tr(\La_i\J^i)\ . \eeq
This takes the place of (\ref{eq:standconslaw}), and of course reduces
to it for $\La_i = 0$, i.e.   $\mu=0$.

\section{Noether theorem for $\mu$-symmetries}

The relation between standard symmetries of the Lagrangian and
conservation laws is described by the classical Noether theorem
\cite{Kos,Noe}. Muriel, Romero and Olver recently gave a
Noether-type theorem for $\la$-symmetries. Here we extend it to
the case of $\mu$-symmetries. Our first result is that when the exact 
invariance of the \LG\ is replaced by a $\mu$-invariance, then the standard 
Noether theorem is 
replaced by a suitably ``corrected'' form. Using the operator $\grad_i$, 
this $\mu$-conservation law  can be  written in the form 
(\ref{eq:muconslaw0}), formally similar to 
a standard conservation law; alternatively, the  r.h.s. of (\ref{eq:nabJ0}) 
can be interpreted as the ``deviation from  the standard conservation law'' 
(i.e. from the case $\mu=0$).

\sk
It is convenient to discuss separately the case of first order
Lagrangians.

\medskip\noindent
{\bf Theorem 1.} {\it Consider the first order Lagrangian
$\L(u,u_x)$ and the vector field $X = \vphi^a (\pa / \pa u^a)$.
Define
\beq\lb{cc1v} (\J^a_{\ b})^i \ := \ \vphi^a \ \pi^i_b \ . \eeq
where $\pi^i_a := (\pa \L / \pa u^a_i)$, then $X$ is a $\mu$-symmetry 
for $\L$ if and only if the {\rm M}-vector
$\J^i$ is a $\mu$-conserved vector.}

\medskip\noindent {\bf Proof.} We observe preliminarily that
for first order Lagrangians, the $\mu$-con\-ser\-vation of $(\J^a_{\
b})^i$ under $X$ (with the notation used above) is
equivalent~to
\beq\label{nabJ}  (\na_i)^a_{\
b}\,(\J^i)^b_{\ a}\=D_i\Tr(\J^i)+(\La_i)^a_{\
b}\,\phi^b\li=
D_i{\bf P}^i+\Tr(\La_i{\cal P}^i) =0 \ ;\eeq
this is easily checked by direct computation.

Let us first show that $\mu$-invariance of the Lagrangian implies
$\mu$-conservation of $\J^a_{\ b}$; we recall that this conservation
can be expressed in the form (\ref{nabJ}). As $X = \vphi^a (\pa / \pa u^a 
)$, its first $\mu$-prolongation is
\beq\label{eq:pr1:1} X^{(1)}_{(\mu)} \ =
\ \vphi^a \, {\pa \over \pa u^a} \ + \ \[ D_i \vphi^a \, + \,
(\La_i)^a_{\ b} \vphi^b \] \, {\pa \over \pa u^a_i} \ . \eeq
Applying this on the Lagrangian $\L$ and integrating by parts, we
get \beq X^{(1)}_{(\mu)} [\L] \ = \ \vphi^a \, \( {\pa \L \over \pa u^a}
\ - \ D_i \, {\pa \L \over \pa u^a_i} \) \ + \ D_i \( \vphi^a
\pi_a^i \) \ + \ (\La_i)^a_{\ b}\, \vphi^b \, \pi_a^i \ ; \eeq the
Euler-Lagrange equations $\E[\L]=0$ grant that the first term
vanishes on solutions to the equations, hence this
reduces to \beq\label{eq:Jevol} X^{(1)}_{(\mu)} [\L] \ = \ \ D_i \(
\vphi^a \, \pi_a^i \) \ + \ (\La_i)^a_{\ b}\, \vphi^b \, \pi_a^i \
. \eeq
Recalling the definition of $\grad_i$, see
(\ref{eq:nabla}), we recognize this as \beq X^{(1)}_{(\mu)} [\L] \ = \
(\grad_i)^a_{\ b} \,    \pi^i_a \, \vphi^b   \ . \eeq
This shows that $X^{(1)}_{(\mu)} [\L] = 0$ implies (\ref{nabJ}).
Going through this computation the other way round, we also show
that $\mu$-conservation of $\J$ implies $X^{(1)}_{(\mu)} [\L] = 0$. \EOP
\bigskip

The formulation of Theorem 1 is limited to first order
Lagrangians. This is often also the case when one quotes the
standard Noether theorem; actually, Noether theory -- i.e. the
relation between symmetries and conservation laws -- also holds
for higher order Lagrangians, albeit with a necessarily more
complex notation\footnote{In view of this fact, our subsequent
results will be stated for first order Lagrangians. Extensions
along the lines of the extensions of Theorem 1 as provided by
Theorem 2 would also be possible; however these would not involve
new ideas, and would instead involve notational heaviness, so that
we avoid to state these.}. The same holds for our generalization,
i.e. Theorem 1 extends to higher order Lagrangians.

We state without proof the following result, which can be verified
using a suitable recursive procedure (see \cite{MRO} for the case
$p=q=1$). Note this extends standard Noether theory via a \quotes{minimal 
substitution}.

\sk\ni {\bf Theorem 2.} {\it Consider the $r$-th order \LG\
$\L=\L(x,u^{(r)})$, and a \vf\ $X$. Then
\begin{itemize}
\item[(a)] $X$ is a $\mu$-symmetry
for $\L$, i.e. $Y[\L]=0$, if and only if there exists {\rm M}-vector
$(\J^i)_{\ a}^{b}$ satisfying the
$\mu$-conservation law \beq\lb{cl} (\na_i)_{\ b}^a \, (\J^i)_{\
a}^{b} \ = \ 0 \ . \eeq
\item[(b)] The {\rm M}-vector $(\J^i)^b_{\
a}$ is obtained in the following way: write the usual current
density vector $({\bf P}_0)^i$ as determined by the given \LG\ and
by the \vf\ $X$ considered as a  standard \sy\ for $\L$ (i.e. with
$\La_i\equiv 0$),  and replace each term
$\dst{\big(D_{J}\phi^a\big)}$ appearing in $({\bf P}_0)^i$ with
$\dst\big({\na_{J} \phi\big)^a}$ ($\,|J|\ge 0$); \item[(c)] the
conservation law (\ref{cl}) is obtained replacing the global
divergence operator $D_i$ with $(\na_i)_{\ a}^b$.
\end{itemize}} \bigskip

This result holds for Lagrangians of any order. For first-order 
Lagrangians the explicit form of the $\mu$-conservation
law was given above, see eq. (\ref{nabJ}) and Theorem 1.
For second order Lagrangians, the $\mu$-conserved vector is
\beq\lb{cc2v} (\J^a_{\ b})^i
\ = \   \phi^b\li + \big((\na_j)^b_{\ c}\, \phi^c\big) \lij -
\phi^bD_j\lij \ . \eeq
It is easy to check directly that  the l.h.s. of the equation
$\nabla_i \J^i\! =\! 0$, in both these cases, i.e. with
$\J$ given by (\ref{cc1v}) and respectively by (\ref{cc2v}), is in
fact equal~to 
\beq Y[\L]+\phi \cdot{\cal E}(\L)\eeq 
where ${\cal E}$ is the Euler-Lagrange operator.

For the vector (\ref{cc2v}) the $\mu$-conservation law (\ref{cl}) can 
be written as $D_i({\bf P}^i)=-(\La_i)^a_{\ b}(\J^i)_{\ a}^b$, as in
(\ref{eq:nabJ0}); alternatively, if one introduces the
matrix-valued vector $\big(\J_0^i\big)_{\ b}^a$ which is obtained putting
$\La_i=0$ in $\big(\J^i\big)_{\ b}^a$, the $\mu$-conservation law
(\ref{cl}) can be also written as \beq D_i({\bf P}_0^i)\=
-D_i\Big(\big(\La_j\phi\big)_a\lij\Big)- (\La_i)_{\
b}^a(\J_0^i)^b_{\ a}-\big(\La_i\La_j\big)_{\ b}^a\, \phi_b\lij
\eeq where the contribution due to the $\La$
matrices is entirely shifted at the r.h.s.

\section{Divergence $\mu$-symmetries}

In the standard Noether theorem, a conserved current can be related, 
as well known,  to a {\it divergence symmetry} of the Lagrangian, i.e. to
a vector field $X = \xi^i (\pa / \pa x^i) + \vphi^a (\pa/\pa u^a)$ such 
that
$X^{(1)} [\L] + (\Div\, \xi) \L = \Div\, B$ (see \cite{Olv1} for
details); dealing with evolutionary representatives, i.e. vertical
vector fields, this reduces to
$X^{(1)} [\L]  =  \Div\, B   \equiv   D_i B^i$.
We say that $X = \vphi^a (\pa / \pa u^a)$ is a {\it divergence
$\mu$-symmetry} for the Lagrangian $\L$ of order $r$ if there
exists a (matrix-valued) $p$-tuple $\B$ such that
\beq\label{eq:divsym} Y [\L] \  = \ \ (\grad_i)^a_{\
b} (\B^i)^b_{\ a} \ = \ {\rm Tr} \[ \grad_i \B^i \] \ . \eeq

\smallskip\noindent
{\bf Theorem 3.} {\it The vector field $X = \vphi^a (\pa / \pa
u^a)$ is a divergence $\mu$-symmetry for the first order
Lagrangian $\L (u,u_x)$ if and only if there is $\B$ such that,
with $\J^a_{\ b} := \vphi^a \pi_b$, $(\J - \B)$ is a
$\mu$-conserved vector.}

\smallskip\noindent
{\bf Proof.} Let us assume that (\ref{eq:divsym}), with $r=1$, is
satisfied. As seen in the proof to Theorem 1, $X^{(1)}_{(\mu)}
[\L] = (\grad^a_{\ b})_i ( \pi_a^i \vphi^b)$; hence
(\ref{eq:divsym}) implies
\beq\label{eq:thm3a} (\grad^a_{\ b})_i \
\( \pi_a^i \, \vphi^b \) \ = \ (\grad^a_{\ b})_i (\B^i)^b_{\ a} \
. \eeq
The term on the left hand side is just $\grad_i \J^i$, and
therefore (\ref{eq:thm3a}) reads
\beq\lb{thm3b} (\grad^a_{\ b})_i \
\[ (\J^i)^b_{\ a} \ - \ (\B^i)^b_{\ a} \]\=0 \ . \eeq
Going through this computation the other way round, we also show
that $\mu$-conservation of $(\J - \B)$ implies $X^{(1)}_{(\mu)} [\L] =
{\rm Tr} (\grad_i \B^i) $. \EOP
\sk

As anticipated, the above result allows us to include in our
discussion the case of evolutionary \vf s and of generalized \vf s
as well.

\smallskip\ni {\bf Theorem 4}. {\it The first-order \LG\ $\L$ is
$\mu$-invariant under the \vf\ $X$ (now possibly with
$\xi^i\not=0$) written in evolutionary form as $X = Q^a (\pa / \pa
u^a)$ if and only if the quantity
$$\big(\J^i-\B^i\big)^b_{\ a}\equiv Q^b\li+\L\,\xi^i\de^b_{\ a}$$
is $\mu$-conserved, i.e. if}
$$\Tr \big(\na^i(\J_i-\B_i)\big)\equiv 
D^i\Jb_i+D^i(\L\,\xi_i)+(\La_iQ)^a\li+
\L\,\Tr(\La_i\,\xi^i)\=0 \ .$$
\ni {\bf Proof.} The condition of $\mu$-invariance for a first order
\LG\ under a generic \vf\ in evolutionary form
$X_Q=Q^a(\pd/\pd u^a)$ can be written 
\beq\lb{evolmu}X^{(1)}_{(\mu)}[\L]\=(\na^i)^a_{\ b}(\B_i)^b_{\ a}
\q {\rm where}\q (\B_i)^b_{\ a}\=-\L\, \xi_i\,\de^b_{\ a} \ ,\eeq
as in (\ref{thm3b}). \EOP
\bigskip

\noindent {\bf Remark 1.} It should be recalled that in the
standard case, divergence symmetries of the Lagrangian are also
symmetries of the corresponding Euler-Lagrange equations. This is
in general {\it not } the case for divergence $\mu$-symmetries, as
already discussed and pointed out in \cite{MRO} for the case of
\ls . See also the subsection \ref{subsect:gaugeq1} below.

\sk \noindent {\bf Remark 2.} We stress that the functional form
of the M-vector $\J$ is the same as that of the standard vector
$\Jb$ met in standard Noether theorem. In particular, if
$\xi^i\equiv 0$, i.e. for $Q = Q (u) = \vphi(u)$, and hence $Q$
not depending on the momenta, $\J$ is homogeneous of order one in
the momenta; allowing generalized symmetries with $Q$ depending on
momenta, the $\J$ would not be homogeneous of order one (we could
thus have conservation laws of higher order in the momenta;
see \cite{LC2,Pucacco} for a discussion of this case and relevant
bibliography) but would however always depend on momenta. Also in
this more general case, all things are completely analogous to
those arising in the same context in the framework of standard
Noether theory and will not be discussed here.

\section{Conditional invariants and $\mu$-symmetries}
\label{sect:conditional}

Special cases of the conservation law (\ref{nabJ}) for
first-order Lagrangians can occur depending on the
form of the r.h.s. of the \eq . For instance, the
case where $\La_i \phi=\la_i \phi$ (with $p>1$) will be considered
in subsections 6.3 and 6.4.

A particularly important  case occurs restricting to a single
(i.e. $p=1$) independent variable; we will then switch to a
``mechanical'' notation, denoting this as $t$ and the dependent
variables as $q^a=q^a(t)$, with $d q^a / dt := \qd^a$. Now $\J$ is
a  single matrix, and the same is for $\La$.

Let us recall that, considering first-order Lagrangians, writing
$p_a:=\pd\L/\pd\qd^a$, if there is some function $\a (q,\qd)$ such
that  $X^{(1)} [\L] = [\a (q,\qd) ] (\vphi^a p_a)$ (recall that
$X^{(1)}$ is the standard first prolongation of $X$), we say that
$X$ is a {\it conditional} symmetry of $\L$: it is indeed a
symmetry of the Lagrangian restricted to the subspace ${\bf P}=
\vphi^a p_a=0$. Then ${\bf P}$ is a {\it conditionally conserved}
quantity, i.e. the relation ${\bf P}=0$ (or, which is the same,
the zero-level manifold for ${\bf P}$) is preserved under the
motion generated by $\L$: in fact, $D_t {\bf P} = X^{(1)} [\L] =
\a {\bf P}$. One speaks then of a {\it conditional invariant}
\cite{Birk,Puc1,SLC}, or also of an {\it invariant relation} in
the sense of Levi-Civita \cite{ALC,LC}. This admits an obvious
interpretation and extension to
$\mu$-symmetries of the Lagrangian.

\medskip\noindent
{\bf Theorem 5.} {\it Let $X=\phi^a (\pa / \pa q^a)$, $\L$ be a
first order Lagrangian, $\mu = \La \d t$ and assume
$\La\phi=\la\phi$ where $\la=\la(t,q,\qd)$ is a scalar function.
With $\J_a^{\ b} := p_a \phi^b$, ${\bf P}=\Tr\,\J$, if one has
$X_{(\mu)}^{(1)} [\L] = \a (q,\qd) {\bf P}$, for some scalar
function $\a$, then
\beq\lb{aml} D_t\,{\bf P}\=(\a-\la)\,{\bf P} \eeq
and ${\bf P}$ is a conditionally $\mu$-invariant quantity.}

\medskip\noindent
{\bf Proof.} The proof follows immediately from the calculations
in Sect.3, which show in particular that
$$ X^{(1)}_{(\mu)}[\L]\=(\na_t)_{\ a}^b(\J)^a_{\ b}\=D_t{\bf 
P}+\la\,\phi^a (\pd\L/\pd
\qd^a)$$ and we reduce to (\ref{aml}) in our hypotheses. \EOP \sk

Alternatively, $X$ may be interpreted as a $\mu$-symmetry for the \LG\
but now with $\la':=\la-\a$. In other words,  ${\bf P}$ is both 
$\mu$-conserved
\big(i.e. $\na_t{\bf P}=(D_t+\la'){\bf P}=0$\big) and conditionally
conserved (i.e. $D_t{\bf P}=0$ on the manifold~${\bf P}=0$).

By the way, the result can be extended to the general case with more
than one independent variable  $(p>1)$. Indeed, coming back to the 
previous notation
$u^a(x^i)$, let us assume that, given a \vf\ $X$ and a form $\mu=\La^i\d 
x_i$,
a  first order \LG\ $\L$ satisfies
$$X^{(1)}_{(\mu)} [\L] \= \Tr\big(A_i\P^i\big)$$
for some $p$ matrices $A_i$: then we  get the conditional
$\mu$-conservation law
$$D_i {\bf P}^i = \Tr \big((A_i - \Lambda_i) \P^i\big) \ .$$

\section{Gauge equivalence and $\mu$-conservation laws}
\label{sect:gauge}

It was shown in \cite{CGM} that given any vector field $X$ in
$\M$, the $\mu$-prolonged vector field $X^{(r)}_{(\mu)}$ in $J^r
\M$ is locally gauge-equivalent (in a sense recalled below) to a
vector field $W$ in $J^r \M$ which is the {\it standard }
prolongation of some vector field $\~X$ in $\M$; moreover, $X$ is
locally gauge-equivalent to $\~X$ via the same gauge
transformation.\footnote{Strictly speaking, this is true for PDEs;
when dealing with ODEs other possibilities appear, as discussed in
\cite{CGM,MuRom1,PuS}, and one can end up dealing with non-local vector
fields of exponential type. In this paper we will however
disregard these possibilities.}

\subsection{Gauge equivalence}
\label{subsect:gaugeequiv}

The gauge transformations mentioned here act in this way: there is
a linear representation $T_g$ of the gauge group\footnote{As
traditional in physical literature, we denote as \quotes{gauge
group} both the finite dimensional group $G$ which acts on the
space $U$, i.e. on the fibers of the bundle $(\M,\pi,\B)$, and the
group of gauge transformations, i.e. of sections $\Gamma$ of the
relevant principal bundle $P_G$ with fiber $G$ \cite{Nak,NaS}; in
this case $P_G$ is a bundle over $\M$ (see \cite{GaeMGS}), and an
element $\ga \in \Gamma$ is a function $\ga (x,u)$ taking values
in $G$.} $G$ acting in the space $U = \R^q$ of the dependent
variables, hence in the fibers of $(\M,\pi,\B)$. This extends to a
representation of $G$ in the fibers of the jet spaces $J^r \M$ of
any order, acting in each of the $U_J \simeq \R^q$ spaces
corresponding to variables $\{ u^1_J , ... , u^q_J \}$ with the
same multi-index $J$ (this was denoted as a {\it jet
representation} in \cite{CGM}).

Then, if $\ga : \M \to G$ is an element of the gauge group, i.e. a
function from $\M$ to the local gauge group $G$, it defines a
matrix-valued function $A_\ga (x,u)$ via 
$ A_\ga (x,u) \=  
T_{\ga (x,u)}$; that is, at $(x,u) \in \M$ we have the
matrix representing the element $g = \ga (x,u) \in G$. The gauge
transformation $\ga$ acts in $U_J$ by these matrices; note that
the action on vector fields will then be described by the
push-forward of this map, given (with $D$ the
differential) by $R_\ga = A_\ga^{-1} (D A_\ga)$.

In the following we will, with an abuse of notation (already used in
\cite{CGM}), simply write $\ga$ for the $R_\ga$ obtained in this way.

Thus, if a vector field $X$ is described in components by 
$X=\phi^a(\pd/\pd u^a)$,
the gauge transformed via $\ga$ will be
\beq
\~X \ = \ \~\phi^a (\pd/\pd u^a) \ = \ (\ga \phi)^a \, (\pd/\pd u^a) \ =
\ \( \ga^a_{\ b} \phi^b \) \, (\pa / \pa u^a ) \ . \eeq

The results recalled above \cite{CGM} guarantee that the vector
field  $\~X$ and its standard prolongation $W$ are gauge
equivalent respectively to $X$ and to its $\mu$-prolongation
$Y=X_{(\mu)}$. We write 
\beq\label{eq:g1}  \~X \= \ga \cdot X \ ,
\ \  W\= \ga \cdot Y \ , \ \ {\rm with}\  \ 
 Y \= \psi^a_J
{\pa \over \pa u^a_J} \ , \ \
 W\=\~\phi^a_J{\pd\ov{\pd u^a_J}}\ ;\eeq
one has then (see \cite{CGM})
\beq\lb{pgf} \psi^a_J\=\g^{-1}\~\phi^a_J \ . \eeq
The setting of $\mu$-symmetries is recovered by writing
\beq\label{eq:g3} \La_i \ = \ \ga^{-1} \, D_i \ga \ ; \eeq in this
case (\ref{eq:hMC}) is automatically satisfied, and conversely
(\ref{eq:hMC}) guarantees (for $q>1$) that locally $\La_i$ can be
written in the form (\ref{eq:g3}) \cite{CGM}.

Suppose now we have a vector field $Y = X^{(r)}_{(\mu)} $ such
that $Y [ \L ] = 0$. That is, $X$ is a $\mu$-symmetry for $\L$ and
hence, by Theorem 1, there is a $\mu$-conservation law associated
to it. One could expect that this implies $\~X[\L]=0$, hence that
$\~X$ would be a standard symmetry of $\L$, and by (standard)
Noether theorem that there would be a standard conservation law
associated to this symmetry. In the next subsections we shall show
that things are not like this in such a generality.

\subsection{The general case}
\label{subsect:gaugegen}

Let us assume  that a \LG\ $\L$ is $\mu$-invariant under some \vf\
$X$, i.e. $X^{(r)}_{(\mu)}[\L]=Y[\L]=0$, and therefore that the
$\mu$-conservation law 
\beq\lb{clrep} \big( \de^a_{\ b} D_i +
(\La_i)^a_{\ b} \big) \, [(\J^i )^a_{\ b} ] \ = \ 0 \eeq 
holds. On the other hand, using (\ref{eq:g3}), the l.h.s. of (\ref{clrep}) is
transformed into 
\beq
\begin{array}{l}  (D_i \J^i )^a_{\ a} \ + \ (\La_i )^a_{\ b} (\J^i)^b_{\ 
a} \ = \
D_i{\bf P}^i+(\gamma^{-1})^a_{\ c}(D_i\gamma)_{\ b}^c(\J^i)^b_{\ a}\ = \\
(\ga^{-1})^a_{\ c}\big[\big(\ga^c_{\ b}D_i+(D^i\gamma)^c_{\ 
b}\big)\,(\J^i)^b_{\ a}\big]\ = \
(\gamma^{-1})^a_{\ c}\,D_i\big(\gamma^c_{\ b}(\J^i)^b_{\ a}\big) \ .
\end{array}\eeq
Thus, the $\mu$-conservation law can be rewritten, in terms of the matrix 
$\gamma$, in the very general and compact form 
\beq
{\tt Tr}\, \big[ \gamma^{-1}D_i\big( \gamma\,\J^i \big)\big] \ = \
0 \ . \eeq 
In particular, this shows that in general no standard
conservation law is associated to the considered \sy .
A significant exception to this conclusion is provided by the case
$q=1$, $p>1$, as discussed below.

\subsection{The case of a scalar field}
\label{subsect:gaugeq1}

In the particular case $q=1$ (that is, a single unknown function
$u=u(x)$ of $p> 1$ variables $\{x^1,...,x^p \}$; in physical terms, a 
scalar field),
the $p$ matrices $\La_i$ are actually $p$ functions $\la_i$, and from the
condition (\ref{eq:comp_cond}) one
deduces the existence (at least locally, see \cite{CGM}) of a
nonvanishing scalar function $\ga=\ga(x,u)$ such that
\beq \lb{gala} \la_i\=  \ga^{-1}\,D_i \ga \ . \eeq

\medskip\noindent
{\bf Theorem 6.} {\it If $q=1,\,p>1$, and a \LG\ is
$\mu$-invariant under the \vf\ $X$, then the conservation law can
be  expressed in the {\rm standard} form \beq\lb{DbfJ} D_i{\bf
\~P}^i\=0 \eeq where ${\bf\~P}^i$ is the ``current density
vector'' determined by the \vf\ $\~X:=\ga X$.}

\sk\ni {\bf Proof.} This easily follows from the above general
formula when there is only one component ($q=1$) for the
considered vectors. More specifically, recalling that if
$X=\phi(\pd/\pd u)$ is a \ms , then there is in this case an
equivalent standard \sy\ $\~X:=\ga X=(\ga\phi) (\pd/\pd u)$, see
\cite{CGM} and the discussion above: one has indeed $\~X^{(r)} =
\ga X^{(r)}_{(\mu)}$ for all orders of prolongations. \EOP

\bigskip

Let us consider in explicit form, for instance, the case of first order
prolongation and first order \LG s.

For the first order $\mu$-prolongation one has
$$X^{(1)}_{(\mu)}\= \phi{\pd\ov{\pd u}}+(D_i\phi+\la_i\,\phi){\pd\ov{\pd
u_i}}\=\ga^{-1}\Big(\ga\,\phi{\pd\ov{\pd
u}}+D_i(\ga\,\phi){\pd\ov{\pd u_i}}\Big)\=\ga^{-1}\~X^{(1)};$$ if
$\L$ is a  first-order Lagrangian, then the $\mu$-invariance
condition reads
$$ \hskip-6.8cm X^{(1)}_{(\mu)}[\L]  =\  \phi{\pd\L\ov{\pd
u}}+\big( D_i\phi+\la_i\phi\big) {\pd\L\ov{\pd u_i}} $$  
$$\hskip 1cm=\
\phi{\pd\L\ov{\pd u}}+\ga^{-1}D_i\Big(\ga\phi{\pd\L\ov{\pd
u_i}}\Big) - \phi D_i {\pd\L\ov{\pd u_i}}   \=
\phi\,\E[\L]+\ga^{-1}D_i\Big(\ga\phi{\pd\L\ov{\pd u^i}}\Big) \= 0
$$
and the conservation law can be given in the
standard form, as stated by theorem~6,
$ D_i{ \bf \~P}^i\equiv D_i\big(\ga\phi\pi^i\big)\= 0 \ .$

\subsection{Conservation laws for multi-component fields}
\label{subsect:gaugeinter}

Theorem 6 holds for $q=1$, and the discussion in subsection
\ref{subsect:gaugegen} shows that for $q>1$ it is in general not
possible to obtain conservation laws in standard form from
$\mu$-symmetries.

Another exception occurs  in this case if, considering a
Lagrangian of order $r$, the matrices $\La_i$ satisfy the
condition
\beq\label{degen} \La_i\,(D_J \phi^a) \ = \ \la_i \,
(D_J \phi^a) \q\q\q \forall J \q {\rm with}\q  \ |J| \le r-1 \ ,
\eeq where $D_J$ is the total derivative with respect to the
multi-index $J$ of order $|J|$, and $\la_i$ are scalar functions.

In this case (mentioned also in subsection
\ref{subsect:intermediate}), a conservation law as in (\ref{DbfJ})
still holds. Rather than discussing the general case, we will
limit to first order Lagrangians; for these ($r=1$) condition
(\ref{degen}) is simply \beq\lb{Lalp} (\La_i \, \phi)^a \ \equiv \
(\La_i)^a_{\ b} \, \phi^b \ = \ \la_i \, \phi^a \ . \eeq

\medskip\noindent
{\bf Theorem 7.} {\it Let $\mu = \La_i \d x^i$ satisfy
equation (\ref{eq:hMC}), and $\L$ be a
first order Lagrangian admitting as $\mu$-symmetry a vector field
$X = \phi^a (\pa / \pa u^a)$ such that (\ref{Lalp}) is satisfied;
let $\ga$ be defined as in (\ref{gala}). Then  the
standard conservation law  holds
\beq D_i \, \( \ga\, \phi^a {\pd \L
\over \pd u_i^a} \) \ = \ 0 \ . \eeq}

\noindent
{\bf Proof.} The argument used in theorem 6 works also in this
case, thanks to  (\ref{Lalp}).~\EOP

\bigskip
\ni Using the same arguments, the result can be  also suitably extended 
to the case of divergence $\mu$-symmetries:

\medskip\noindent
{\bf Theorem 8.} {\it Assume either $q=1,\ p>1$; or that
(\ref{Lalp}) is satisfied if $q>1$. Let $\L$ be a first-order \LG,
$X$ a \vf , $\mu = \La_i \d x^i$ a form satisfying  the equation
(\ref{eq:hMC}), with $\ga$ defined as in (\ref{gala})
and ${\bf P}^i = \phi^a \pi_a^i$ the usual current
density vector.
\begin{itemize}
\item[(a)] If there is a $p-$tuple $B^i$ such that $
X^{(1)}_{(\mu)}[\L] = \ga^{-1} \, \Div\ B$, then the conservation
law in standard form $ D_i ( \ga {\bf P}^i - B^i) =  0$ holds;
\item[(b)] If there is a $p-$tuple $B^i$ such that
$X^{(1)}_{(\mu)}[\L] = \la_i B^i + \Div \, B $, then the
conservation law in standard form $D_i ( \ga {\bf P}^i - \ga B^i)
= 0$ holds. \end{itemize} }

Let us finally remark that clearly if $X$ is a $\mu$-\sy\ (or also
a divergence $\mu$-\sy ) for a Lagrangian, then the corresponding
Euler-Lagrange equation turns out in this case to be symmetric
with respect to the gauge-equivalent standard \sy\ $\~X=\ga X$.

\section{Variational problems with $\mu$-prolonged\\ variation field.}
\label{sect:muEL}

In this section we reverse our point of view, by the introduction of the 
notion of {\it modified} $\mu$-Euler-Lagrange equations.

The Euler-Lagrange equations are indeed obtained requiring that the
action integral $S = \int \L\,  \d^p x$ is stationary under a
variation $\de u$ (vanishing on the frontier $\pa B$ of the region
of integration) of the dependent variables $u^a (x)$. The
variation field $V = (\de u)^a (\pa / \pa u^a)$ induces a
variation of the $u^a_i$ as well; standard Euler-Lagrange
equations (that we have considered so far) are obtained in the
case where this variation is described by the standard lift of the
variation $\de u$, i.e. by the standard prolongation of $V$.

We will now consider the case where the variation
field is $\mu$-prolonged to obtain the variation in the space of
field derivatives.

\subsection{Derivation of the $\mu$-Euler-Lagrange equations}

Let us consider a first order Lagrangian $\L (u,u_x)$, and the
action integral
$$ S(A) \ = \ \int_A \,  \L (u,u_x) \ \d^p x \ . $$
We consider the variation $u \to u + \de u$
corresponding to the action of a vector field $V = \eta^a (x,u)
(\pa / \pa u^a)$; that is, $ \de u^a = \eps \eta^a (x,u)$;
needless to say this also acts on the $u^a_i$.

We consider the case where this action on field derivatives is
described by the $\mu$-prolongation of $V$, i.e. by
$$ V^{(1)}_\mu \ = \  \eta^a \, {\pa \over \pa u^a} \ + \ \[ D_i
\eta^a \, + \, (\La_i)^a_{\ b} \eta^b \] \, {\pa \over \pa u^a_i}
\ ; $$ 
this corresponds to
\beq\lb{A3} \de u^a \ = \ \eps \, \eta^a \ ; \ \
\de u^a_i \ = \ \eps \, (\grad_i)^a_{\ b} \, \eta^b \ . \eeq
The equations
satisfied by fields $\overline{u}$ such that the variation $(\de
S(A))[u]$ of $S$ under (\ref{A3}) vanishes at $u = \overline{u}$ for
all variations $\de u$ satisfying $[\de u]_{\pa A} = 0$, will be
called the {\it $\mu$-Euler-Lagrange equations} and are derived in the
same way as the standard Euler-Lagrange equations.

The variation of $S$ under (\ref{A3}) is given by
\beq\lb{A4} \de S \ = \ \int_A (\de L) \, \d^p x  \ = \
\int_A \[ {\pa \L \over \pa u^a} \, \de u^a \ + \ {\pa \L \over \pa
u^a_i} \, \de u^a_i \] \, \d^p x  \ . \eeq 
According to
(\ref{A3}), $\de u^a_i =  D_i (\de u^a)   +   (\La_i)^a_{\ b} \,
\de u^b$; substituting this into (\ref{A4}) and
integrating by parts, we have
$$ \de S \, = \, \int_{\pa A} {\pa \L \over \pa u^a_i} \,
\de u^a \, \d \sigma \, + \,
\int_A \[ \( {\pa \L \over \pa u^a} - D_i {\pa \L \over \pa
u^a_i} \) \de u^a \, + \, {\pa \L \over \pa u^a_i} (\La_i)^a_{\
b} \de u^b \] \, \d^p x  $$
where $\d \sigma$ is the (surface) element of integration on $\pa A$.
The boundary term vanishes (due to $\de u = 0$ on $\pa A$), and 
rearranging indices we get hence
$$ \de S \ = \ \int_A  \[ {\pa \L \over \pa u^a} \ - \
D_i {\pa \L \over \pa u^a_i}  \ +
\ {\pa \L \over \pa u^b_i} (\La_i)^b_{\ a} \] \ \( \de u^a \) \ \d^p x \ . 
$$
As we require this to be zero for any choice of $\de u^a$, the term in 
square
brackets has to vanish, i.e. we have established the $\mu$-Euler-Lagrange
equations to be
\beq\lb{muEL} {\pa \L \over \pa u^a} \ - \ D_i {\pa \L \over \pa
u^a_i}  \ + \ {\pa \L \over \pa u^b_i} (\La_i)^b_{\ a} \ = \ 0 \ .
\eeq

\sk\ni {\bf Remark 3.} Writing $\La^T_i$ for the transpose of $\La_i$, 
and (as above)
$\pi_a^i := (\pa \L / \pa u^a_i)$ for the momenta,
the (\ref{muEL}) are also written as
$$ (\pa \L / \pa u^a) \ - \ D_i \, \pi_a^i \ = \ - \
(\La^T_i)_{a}^{\ b} \ \pi_b^i  \ . $$

\sk\ni {\bf Remark 4.} It would not be difficult, proceeding along
these lines in the same way as in the standard case (i.e. by
repeated integration by parts), to obtain the $\mu$-Euler-Lagrange
equations for higher order Lagrangians.

\subsection{Conservation laws from $\mu$-symmetries}

We will now consider the case where $\L$ admits a $\mu$-symmetry
$X = \phi^a (\pa / \pa u^a)$, with $\mu$ the same form appearing
in the $\mu$-prolongation of the variation field; we will as usual
denote by $Y$ its $\mu$-prolongation, $Y = \phi^a (\pa / \pa u^a)
+ [ (\grad_i)^a_{\ b} \phi^b ] (\pa / \pa u^a_i)$, so that our
assumption is that $Y [\L ] = 0$. Recalling also $(\grad_i)^a_{\
b} = \de^a_b D_i + (\La_i)^a_{\ b}$, the symmetry condition
reads\footnote{This is of course the same as the condition met
above, see e.g. Theorem 1: the symmetry vector field and the
Lagrangian are the same as in the standard variation case, the
difference residing only in considering the (standard or instead
$\mu$-) Euler-Lagrange equation.} \beq\lb{ysym} \phi^a {\pa \L
\over \pa u^a} \ + \ \( D_i \phi^a + (\La_i)^a_{\ b} \phi^b \)
{\pa \L \over \pa u^a_i} \ = \ 0 \ . \eeq

\medskip\noindent
{\bf Theorem 9.} {\it Let $\L$ be a first order \LG, admitting the
vector field $X = \phi^a (\pa / \pa u^a)$ as a $\mu$-symmetry for
a certain form $\mu$. Then the vector ${\bf P}$ of components ${\bf P}^i
= \phi^a \pi^i_a$ defines a standard conservation law, $D_i {\bf P}^i =
0$, for the flow of the associated $\mu$-Euler-Lagrange
equations.}

\medskip\noindent
{\bf Proof.} With an integration by parts, (\ref{ysym}) is
transformed into
$$ \phi^a {\pa \L \over \pa u^a} \ + \  D_i \(
\phi^a  {\pa \L \over \pa u^a_i} \) \ - \ \phi^a D_i \(  {\pa \L
\over \pa u^a_i} \) \ + \ \( (\La_i)^b_{\ a} \phi^a \) {\pa \L
\over \pa u^b_i} \ = \ 0 \ . $$ (We have exchanged indices in the
last term -- which is fine since they are both summation indices.)
Collecting the terms with $\phi^a$, we get this is in the form
$$ \phi^a \, \[ {\pa \L \over \pa u^a} \, - \, D_i \(  {\pa \L
\over \pa u^a_i} \) \, + \, {\pa \L \over \pa u^b_i} (\La_i)^b_{\
a} \] \ + \  D_i \( \phi^a  {\pa \L \over \pa u^a_i} \) \ = \ 0 \ . $$
The term in square brackets vanishes due to
(\ref{muEL}), hence (\ref{ysym}) implies the conservation law
\beq\lb{A13} D_i \( \phi^a  {\pa \L \over \pa u^a_i} \) \ = \ 0 \
, \eeq
as claimed in the statement. \EOP

\subsection{Discussion}

We stress that (\ref{A13}) is a standard conservation law, whose
conserved vector ${\bf P}$ has the familiar form of conserved vector
for the standard Noether theorem. Hence a $\mu$-symmetry is for the
$\mu$-variational equations what a standard symmetry is for the standard
variational equations.

This should not be surprising, in view of the discussion in
section \ref{sect:gauge}. In fact, now both the variation vector
fields  $V_{(\mu)}$ and the symmetry vector field $Y = X_{(\mu)}$
are $\mu$-prolongations (with the same $\mu$) of Lie-point vector
fields in $\M$, and according to \cite{CGM} they could be taken to
be standard prolongations of (different) Lie-point vector fields
via a gauge transformation. With reference to the arbitrariness of
the variation vector field, it is worth noting that any gauge
transformation would transform vector fields in $\M$ vanishing on
a set $\pa A \ss \M$ into (generally, different) vector fields
also vanishing on the same set.

Thus, the case of $\mu$-symmetries for a Lagrangian $\L$ and
evolution law  described by $\mu$-Euler-Lagrange equations for
$\L$ should be seen (locally, see \cite{CGM}) as a gauge
transformation of the situation where $\L$ has a standard symmetry
and the evolution law is given by the standard Euler-Lagrange
equations.

As suggested by this correspondence, the situation where we have
an evolution controlled by the $\mu$-Euler-Lagrange equations
(\ref{muEL}) and a standard symmetry is symmetric (or, maybe more
precisely, dual) to the situation studied in sections 3-6 above,
and would produce a $\widetilde{\mu}$-conservation law (we use the
notation $\widetilde{\mu}$ to stress this is a $\mu$-conservation
law with a different $\mu$; actually: $\widetilde{\mu} = -
\mu$).

In fact, consider the $\mu$-Euler-Lagrange equations (\ref{muEL})
for a Lagrangian $\L$ and a form $\mu = \mu_0 = \La_i \d x^i$; and
assume this admits the vector field $X = \phi^a (\pa / \pa u^a)$
as a standard symmetry, so that $X^{(1)} [\L] = 0$. Recalling the
(standard) prolongation law, this means
$$ \phi^a \( {\pa \L \over \pa u^a} \) + (D_i \phi^a ) \( {\pa \L \over 
\pa u^a_i} \) \
\equiv \ \phi^a \[ {\pa \L \over \pa u^a}  - D_i \( {\pa \L \over \pa 
u^a_i} \) \] +
D_i (\phi^a \pi_a^i ) \ = \ 0 \ . $$
Using (\ref{muEL}), this reads
$$ D_i (\phi^a \pi_a^i ) \ = \ {\pa \L \over \pa u^b_i} \, (\La_i )^b_{\ 
a} \, \phi^a \ , $$
which is indeed a $\mu$-conservation law, see section 2, for $\mu
= - \La_i \d x^i = - \mu_0$.

\section{Applications and Examples}

We will now discuss briefly some general aspects of $\mu$-\sys\ and 
$\mu$-conser\-vation laws, together with some of their possible
interpretation and applications.

First of all, as already pointed out, one of the main properties coming 
from the presence of a $\mu$-\sy\ of a given \LG\ is that the ordinary 
Noether theorem 
is replaced by a suitable modification of the conservation law.
It is clear, on the other hand,  that the existence of a 
conservation law (even in a weakened or modified form) can be {\it per
se} a relevant result.

More specifically, in some cases, the $\mu$-\sy\
can be obtained as a perturbation of an exact \sy , and --
accordingly -- the $\mu$-conservation as a perturbation of the
standard Noether law. This will be illustrated by some examples
below, see in particular examples 1,5 and 6. Example 1 will be also 
considered in the setting of section 7, i.e. by the introduction of the
$\mu$-modified  Euler-Lagrange \eq s.

More generally, consider a -- say, first order -- Lagrangian
$\L_0$ (and/or second order equations $\E_0$, possibly the
Euler-Lagrangian equations for $\L_0$) which are symmetric under a
vector field $X$, i.e. under its first (respectively, second)
standard prolongation.
Consider now a perturbation $\L_\eps = \L_0 + \eps \L_1 + \eps^2
\L_2 + ... $ (and/or $\E_\eps = \E_0 + \eps \E_1 + \eps^2 \E_2 +
...$); we wonder if this may admit $X$ as a $\mu$-symmetry with
nontrivial $\mu$ (as we assumed $X$ is not a standard symmetry of
$\L_\eps$ or $\E_\eps$).

We have shown that in some cases the  $\mu$-conservation law
actually reduces to a standard one, and also pointed out the existence
of a relation with conditional invariants: see examples 4,6 and 7.

Another important aspect related to the presence 
of $\mu$-\sys\ is the possibility 
of reducing the order of differential \eq s.
In the case of variational problems, the first difficulty   -- as
well known \cite{MRO} -- is that the Euler-Lagrange \eq s do not
inherit, in general, the $\mu$-invariance of the \LG ; this is
clearly a strong difference with respect to the case of exact \sys .
As a consequence, only a ``partial'' reduction can be expected,
along the same lines pointed out in \cite{MRO}. Some aspects of 
this nontrivial problem will be illustrated in examples 3,4 and 7 below.

One of the differences between standard conservation law and a 
$\mu$-conservation law lies in that a standard conservation law allows for 
an algebraic substitution eliminating one of the momenta (or variables) 
in the Euler-Lagrange \eq s, whereas a $\mu$-conservation law amounts 
to an auxiliary 
differential \eq\  which has to be solved in order to implement the 
reduction allowed by 
the conservation law itself. This will be useful if the auxiliary \eq\ is 
easier to solve (in general, or at least in order to provide some 
special \so ) than the Euler-Lagrange \eq s.

In some cases, as shown in examples 8 and 9, it can be possible
to explicitly integrate the 
$\mu$-conservation law: this may help to obtain solutions, or at least 
to suggest how to find particular solutions
to the Euler-Lagrangian equations. 

We finally refer to \cite{GaeMGS} for a different interpretation of
$\mu$-\sys\ in  terms of changes of frame, or better of  (pointwise
linear) transformations  acting  in the same way on the vertical vector
spaces for the fibrations 
$J^{k+1}M\to J^k M$ ($k\ge 0$).

We are now proposing some examples illustrating the above discussion;
some of the examples are, admittedly, rather artificial, but they can be
useful to clarify also more technical details of the procedure.

In the examples below, when  the \LG\ involves two
functions  depending on two variables, we shall write, to simplify
notations,  $x,y$ instead of $x^1,\, x^2$, and $u=u(x,y), \,
v=v(x,y)$ instead of $u^1,\,u^2$. If there is only one independent
variable, we will use the same ``mechanical''  notation $q^a(t)$ as  in
section 5 (but, in order to avoid any confusion with powers, we will
actually adopt  the notation with lower indices $q_1(t),\,
q_2(t)$).

\medskip

\sk\ni {\bf Example 1.}  Consider the \vf
\beq\lb{ex1}X\=u{\pd\ov{\pd u}}+{\pd\ov{\pd v}}\eeq
where $u=u(x,y),\, v=v(x,y)$ and
with the form $\mu = \La_1 \d x + \La_2 \d y$ defined by
$$\La_1\=\pmatrix{0 & 0\cr u_x  & 0} \q\q \La_2\=\pmatrix{0 & 0\cr u_y  &
0} \ . $$ 
It is easy to check that the \LG
$$\L\= \unm\Big( u_x^2+u_y^2\Big)-{1\ov
u} \big( u_xv_x+u_yv_y \big) + u^2 \exp(-2v) $$ is $\mu$-invariant
(but not invariant)   under the \vf\ $X$. 
The M-vector $(\P^i)^b_{\ a}$ is explicitly given by
$$\big((\P^1)^b_{\ a}\, ,\, (\P^2)^b_{\ a}\big) \= \(
\pmatrix{uu_x-v_x & -u_x\cr u_x-v_x/v & -u_x/u}\, ,\, 
\pmatrix{uu_y-v_y & -u_y\cr u_y-v_y/v & -u_y/u}\)\ ;$$
it satisfies the $\mu$-conservation law $\Tr (\na_i \J^i)=0$, which takes 
here the form
$$D_i {\bf P}^i\equiv D_x\(uu_x-v_x-{u_x\ov{u}}\)+
D_y\(uu_y-v_y-{u_y\ov{u}}\)\=  u_x^2+u_y^2$$
where ${\bf P}^i=\Tr(\P^i)$, using also the Euler-Lagrange \eq s, or
$$u(1-u^2)(u_{xx}+u_{yy})+u^2(v_{xx}+v_{yy})\= u_x^2+u_y^2 \ .$$
In agreement with Theorem 1, the r.h.s. of  this expression is precisely
equal to
$\dst{-\Tr(\La_i{\P}^i)=-(\La_i\phi)^a\li}$. 
Notice that the quantity $u_x^2+u_y^2$ is just the ``\sy -breaking term", 
i.e. the term which prevents the above \LG\ from being exactly symmetric
under  the \vf\ (\ref{ex1}). 

Similar results hold  for any  \LG\ depending, more in general,  on the 
quantities
$x,\, y,\, z_1\!\!=\! e^{- v} u \, ,  \, z_2\!\! =\! e^{- v} u_x \, ,   
\, w_1\!\! =\! u_x^2/2 - (u_x v_x)/u \, , \,   w_2\!\! = u_y^2/2 -~
(u_y v_y)/u$. 

\sk\ni
{\bf Example 2.}  Let us
consider the same setting and notation of Example 1 above, but we adopt 
here the point of view of sect.7. The modified $\mu$-Euler-Lagrange 
equations (\ref{muEL}), for a generic first order \LG , read 
\beq\lb{eq:ex71} 
D_x \( {\pa \L \over \pa u_x} \) + D_y \( {\pa \L \over \pa u_y} \) \ = \
{\pa \L \over \pa u} + u_x {\pa \L \over \pa v_x} + u_y {\pa \L \over \pa 
v_y} \eeq  
\beq\lb{eq:ex72} \hskip-2.8cm D_x \( {\pa \L \over \pa v_x} \) + D_y  \,
\( {\pa \L \over \pa v_y} \) \ = \
{\pa \L \over \pa v} \ .  \eeq
In the present case 
${\bf P}$ is the vector of components
$$ {\bf P} \=   \( u {\pa \L \over \pa u_x} + {\pa \L 
\over \pa v_x} \ ,
\ u {\pa \L \over \pa u_y} + {\pa \L \over \pa v_y} \) $$
and the conservation law claimed by Theorem 9 will be given by the 
vanishing~of
$$ 
D_i {\bf P}^i \=    u_x   {\pa \L \over \pa u_x}  + u D_x   
{\pa \L \over \pa u_x}   + D_x   {\pa \L \over \pa v_x}     
 +   u_y   {\pa \L \over \pa u_y}  + u D_y   {\pa \L \over 
\pa u_y}   + D_y   {\pa \L \over \pa v_y}     $$   
$$ =  u \( {\pa \L \over \pa u} + u_x
{\pa \L \over \pa v_x} + u_y {\pa \L \over \pa v_y} \) + {\pa \L
\over \pa v}
 + u_x {\pa \L \over \pa u_x} + u_y {\pa \L \over \pa
u_y}$$
having used  (\ref{eq:ex71}-\ref{eq:ex72}). Let us now assume $\L$ 
admits $X$ as a $\mu$-symmetry:   as said  in Example 1 this is the case
if $\L = \L (x,y,z_1,z_2,w_1,w_2 )$. It is easy to conclude, 
by direct calculation, that
indeed $D_i {\bf P}^i \equiv 0$, as claimed by Theorem~9.

\sk\ni
{\bf Example 3.} 
Consider the \vf , with $\xi_i\not=0$,
$$X\=x{\pd\ov{\pd x}}+{\pd\ov{\pd v}}$$
or, in evolutionary form,
$$X_Q\= -xu_x{\pd\ov{\pd u}}+(1-xv_x){\pd\ov{\pd v}}$$
and the $\mu$-form defined by the two matrices
$$\La_1\=\La_2\=\pmatrix{0 & 0\cr 1 & 0} \ .$$
A \LG\ satisfying the $\mu$-invariance condition (\ref{evolmu})
is, for instance,
$$ \L\= {1\ov x}\Big(u^2+x^2u_x^2+u_y^2+(xvu_x+v_y)^2\Big) \ .$$
The vector ${\bf P}$ is
$${\bf P}\equiv\( -xu_x{\pd\L\ov{\pd u_x}},\ -xu_x{\pd\L\ov{\pd 
u_y}}+(1-xv_x){\pd\L\ov{\pd v_y}}\)\ .$$
In the case of {\it exact} invariance of $\L$ under $X$, one would expect 
the conservation law
$D_x{\bf P}^1+D_y{\bf P}^2+D_x(x\L)\=0$;
in our case, the $\mu$-invariance of $\L$ produces the following modified 
conservation law, in agreement with Theorem 4,
$$D_x{\bf P}^1+D_y{\bf P}^2+D_x(x\L)\=-(\La_iQ)^a{\pd\L\ov{\pd u^a_i}}=
xu_x{\pd\L\ov{\pd v_y}}\=2(xvu_x^2+u_xv_y) \ .$$
It could be perhaps more interesting to look for the \sy\ properties of 
the Euler-Lagrange \eq s for this \LG . Let us recall indeed that the 
$\mu$-invarian\-ce of the \LG\ is in general not shared by the 
Euler-Lagrange 
\eq s; however,  this does not exclude that one at least of these 
\eq s can be expressed in terms of $\mu$-invariant 
quantities (and of derivatives thereof). This is precisely our case:
indeed,  introducing the quantities
$\a\=xu_x$, $\be\=xvu_x+v_y$,
which are $\mu$-invariant, i.e. $X_{(\mu)}^{(1)}(\a)=
X_{(\mu)}^{(1)}(\be)=0$, one has that one of the Euler-Lagrange \eq s 
can be written
$$\be_y-\a\be\=0\ .$$
This property is clearly of great help in finding explicit \so s; e.g., 
choosing the simplest possibility $\be=0$, it is easy to obtain for 
instance the particular \so\ $u=(\log x)\exp(y),\ v=\exp(-e^y)$ of 
the Euler-Lagrange \eq s.

\sk\ni 

\sk\ni {\bf Example 4.}  We now consider a time-dependent problem 
for the two variables $q_1(t),\, q_2(t)$; let
$$ X \= q_1{\pd\ov{\pd q_1}}+{\pd\ov{\pd q_2}} $$
and the $\mu$-form $\mu = \La \d t$, where $\La=\la I$
with $\la=q_1$. A \LG\ $\mu$-invariant under $X$ is for instance
$$ \L \= \unm \Big( {\qd_1\ov{q_1}}-q_1\Big)^2 + \unm(\qd_2-q_1)^2 $$
and it is now easy to verify  the
existence of the $\mu$-conservation law $\na_t{\bf P}\=0$ for 
the quantity ${\bf P}=(\qd_1/q_1)+\qd_2-2q_1$. This
can be also interpreted as a conditional invariant relation 
(cf. sect. 5) in the form
$ D_t{\Jb} =  X^{(1)}_\mu(\L)-\la{\bf P} \= -q_1 {\bf P} \ .$
If we now introduce the $\mu$-invariant quantities 
$\a=(\.q_1/q_1)-q_1$ , $\be=\.q_2-q_1$, 
it is easy to see that the Euler-Lagrange \eq s (in this case a system 
of ODE's) can be written in the form
$$\a_t +q_1(\a+\be)=0\q ,\q \be_t=0\ .$$ 
Then, as expected, they are not invariant under $X$, nor 
$\mu$-invariant (or, better, only the second one of them is 
$\mu$-invariant). 
However, this is a good example of a ``partial'' reduction, in 
the same sense as explained in \cite{MRO}: these \eq s indeed admit a
reduction into  the system of {\it first-order} \eq s $\a=\be=0$.  This
of course allows to obtain  easily the particular \so\
$q_1=c_1/(1-~c_1t)\, , \, q_2=c_2-\log(1-c_1t)$,  where $c_1,c_2$ are
arbitrary constants.

This also  generalizes to a system of \eq s
the result given in the above mentioned paper \cite{MRO}.

\sk \ni{\bf Example 5.} We consider here a problem for
the two variables $q_1(t),\, q_2(t)$ admitting as a $\mu-$\sy\ the
rotation operator
$$ X\=q_2{\pd\ov{\pd q_1}}-q_1{\pd\ov{\pd q_2}} \={\pd\ov{\pd\th}} $$
having introduced as new variables the polar coordinates $r=r(t)
, \th=\th(t)$. Let us assume that $\mu$ is given by $\mu=\La \d t$
where, in the basis $r,\,\th$,
$$\La\=\pmatrix{0 & 0 \cr 0 & \eps\cos\th}$$
and $\eps$ is a ``small'' real parameter. A $\mu-$invariant \LG\ is, e.g.,
$$\L\=\unm\big(\.r^2+r^2(\.\th-\eps\sin\th)^2\big)-V(r)$$
where $V(r)$ is an arbitrary function. This \LG\ can be viewed as
a ``perturbation'' (for $\eps\not=0$) of a rotationally-symmetric
\LG . It is simple to verify indeed that the angular momentum 
$r^2\.\th$ is not conserved; we have instead 
the $\mu$-conservation law  
$D_t{\bf P}\= (-\eps\cos\th)\,{\bf P}$ for the quantity 
$${\bf P}\= {\pd\L\ov{\pd\.\th}}\=r^2(\.\th-\eps\sin\th) \ .$$

\sk\ni
{\bf Example 6.} As in the previous example, we consider the
rotation \vf\ $X=\pd/\pd\th$ as a (variational) $\mu$-\sy , but here the
variables $r, \th$ are the independent variables and a single
dependent variable $u=u(r,\th)$ is introduced; then $\phi=0\, ,\, 
\xi_1=0\, , \, \xi_2=1$.  We consider the \LG
$$\L\= \unm r^2\exp(-\eps\th)u_r^2+\unm \exp(\eps\th)u_\th^2$$
which is clearly not invariant under rotation \sy\ (if
$\eps\not=0$), but is $\mu$-invariant with $\mu=\eps\d \th$.
The above \LG\ is the \LG\ of a perturbed Laplace \eq , indeed the 
Euler-Lagrange \eq\ is the PDE
$$r^2u_{rr}+2ru_r+\exp(2\eps\th)(u_{\th\th}+\eps\, u_\th)\=0 \ .$$
It is easy to check that the current density vector
$${\bf P}\equiv\(-r^2\exp(-\eps\th)u_ru_\th\, , \, \unm 
r^2\exp(-\eps\th)u_r^2-\unm \exp(\eps\th)u_\th^2\)$$
satisfies the $\mu$-conservation law
$$D_i{\bf P}_i\=-\eps {\bf P}_2 \ .$$
In agreement with Theorem 6, in this case also the (standard) 
conservation law $D_i\~{{\bf P}}^i=0$ holds, with 
$$\~{{\bf P}}\equiv \Big(-r^2 u_ru_\th\, , \, \unm r^2 u_r^2-\unm 
\exp(2\eps\th)u_\th^2\Big)$$
thanks to the (local, see  \cite{CGM}) equivalence to the standard \sy\
\vf\ $\~X\=\gamma\,X\=\exp(\eps\th)\,{\pd/\pd\th}$.

\medskip\ni {\bf Example 7.} 
This is another modification of the \LG\ of the Laplace \eq . The 
following \LG\ for the function $u=u(x,y)$
$$\L\=\unm \exp(2u)\(u_x^2+u_y^2\)+{1\ov{3}}\exp(3u)u_y^3$$
is clearly not invariant under $X=\pd/\pd u$, but it turns to be 
$\mu$-invariant defining
$\mu=-u_x\d x-u_y\d y $.
The Euler-Lagrange \eq\ is the PDE
$$u_{xx}+u_{yy}+u_x^2+u_y^2+2\exp(u)u_y(u_y^2+u_{yy}) \=0 \ .$$
Rather than the expression of the $\mu$-conserved current, it can be 
interesting in this case to remark that, according to the discussion in 
subsect.6.3, there is, as in the above example, an equivalent exact 
\sy\ $\~X=\exp(-u)X$ for this \LG ; as a consequence, the 
Euler-Lagrange \eq\ turns out to be  
{\it both} exactly symmetric under $\~X$ and $\mu$-symmetric under $X$. 
Indeed, introducing the quantities
$\a\=u_x\exp(u)$, $\be=u_y\exp(u)$,
which are invariant under $\~X$ (and $\mu$-invariant under $X$), the 
Euler-Lagrange \eq\ takes  the  invariant form
$$\a_x+\be_y+2\be\be_y\=0\ .$$
Clearly, now, all usual procedures for symmetric \eq s can be applied.

\medskip\noindent
{\bf Example 8.}
The Lagrangian for the two variables $q_1(t),\, q_2(t)$
$${\cal L}\ =\ {1\over {2}}\dot q_1^2+{1\over {2}}q_1^2(\dot q_2-q_2)^2$$
is $\mu$-invariant under
$$X\ =\ {\partial\over{\partial q_2}}\quad {\rm with}\quad 
\Lambda=\pmatrix{0 & 0\cr  0 & 1}\ .$$
The $\mu$-conservation law is $$D_t{\bf P}=-{\bf P} \ ,$$
where ${\bf P}\=  q_1^2(\dot q_2-q_2)$.
Integrating this $\mu$-conservation law one gets
$$ {\bf P}\ =\ q_1^2(\dot q_2-q_2)\ =\ c_1\exp(-t) \quad (c_1={\rm const})$$
and, from the Euler-Lagrange equations, one gets the other equation
$$q_1^3\ddot q_1\ =\ c_1^2\exp(-2t) \ .$$
It is now easy to obtain the solution
$$q_1(t)\ =\ \sqrt{2c_1}\,\exp(-t/2)\ , \ q_2(t)\ =\ c_2\,\exp(t)-\unm\ .$$

\medskip\noindent
{\bf Example 9.}
The following Lagrangian for $u=u(x,y),\,v(x,y)$
$${\cal L}\ =\ {1\over 2}v^2(u_x-u)^2+{1\over 2}(u_y-u)^2+v_xv_y$$
is $\mu$-invariant under
$$X\ =\ {\partial \over{\partial u}}$$
with $\Lambda_1=\Lambda_2=I$. The $\mu$-conservation law is 
$$D_x{\bf P}_1+D_y{\bf P}_2\ =\ -{\bf P}_1-{\bf P}_2$$
where 
$${\bf P}\ =\ \big(v^2(u_x-u)\, ,\, u_y-u\big) \ .$$
One can look for a possible solution imposing e.g. ${\bf P}_1=0$, and 
then 
${\bf P}_2=A(x)\exp(-y)$ where $A(x)$ is a function to be determined. 
It is now 
simple to verify that the Euler-Lagrange equations  are solved by 
$u=\exp(x)\big(c_1\exp(y)+c_2\exp(-y)\big)$ and $v$ any solution to the 
\eq\ $v_{xy}=0$. Putting instead ${\bf P}_2=0$, then a solution is given 
e.g. by 
$\dst{u=-\exp(y),\, v=\exp\(-x/2\)\,\exp\(-e^{2y}/2\)}$.

\section{Final remarks}

We have presented several aspects and different points of view
concerning $\mu$-symmetries and their relationships with
$\mu$-conservation laws and with standard symmetries as well. 

The notion of ``conservation law'' is a classical theme of theoretical 
and mathematical physics which certainly  does not need to be emphasized 
here (see e.g.
\cite{Kos,KrV,LLmec,LLft,Olv1,CaS}). Several attempts are also present 
in the literature towards generalizations (or weakened forms) of this 
notion, 
and we would like to quote in particular \cite{ABnlx} (see also 
\cite{Ax,ABx,AD,ABT} and references
therein) for physical applications, where attention is focused on nonlocal 
conservation laws. 

The general
relevance of the Lagrangian approach is also witnessed by the efforts
made by various authors in order to embed in a variational context
also generic (non-variational) problems (see e.g. \cite{Ibx,Ib2}).

It should be mentioned, in connection with \cite{ABnlx}, that a nontrivial 
relation between $\la$- and $\mu$-\sys\ on the one hand and nonlocal \sys\ 
on the other has been determined \cite{DCF}; this suggests there could be 
relations between our present results and those in \cite{ABnlx}.

Last but not least,  we like to mention that the results given here 
have been reinterpreted by P. Morando in terms of standard conservation
laws under a  deformed Lie derivative, coherently with her geometrical
approach to 
$\mu$-prolongations and \sys\ \cite{PMo}. 

\bigskip\bigskip


\end{document}